\documentclass[10pt,a4paper]{article}  %% \pdf=12

\newcommand\nc{\newcommand}  \nc\rc{\renewcommand}  %% !!!!
\nc\VOMIT[1]{#1}  \nc\OMIT[1]{}  %% swithcing on/off manuscript parts.

\usepackage{graphicx}
\usepackage{float}  %% !!!!

\usepackage{tabularx}
\usepackage{diagbox}
\usepackage{booktabs}

\usepackage{color}
\usepackage{colortbl}
\definecolor{gray}{gray}{0.5}
\nc\gc[1]{\cellcolor{white} #1}
\nc\G[3]{{\color{gray}#1^#2_#3}}  %% !!!!
\usepackage{amsmath}
\usepackage{siunitx}
\nc\m[1]{$ #1 $}
\nc\re[1]{(\ref{#1})}
\nc\F[2]{#1/#2}
\nc\pdv[2]{\frac{\partial #1}{\partial #2}}
\nc\Pdv[2]{\F{\partial #1}{\partial #2}}
\nc\RR[1]{\stackrel{\re{#1}}{=}}
\nc\0[2]{\mathopen{}\mathclose{\9#1{#2}}}  %% \9 without extra spaces around
\nc\1[2]{\Delim#1#2\deliM#1}               %% \11{x} = (x)  etc., also for text
\nc\9[2]{\left\DElim#1#2\right\delIM#1}    %% \91{x} = \left( x \right)  etc.
\nc\Delim[1]{\ifcase #1\or(\or[\or\{\or\mathord<\or\langle\or|\or\|\fi}
\nc\deliM[1]{\ifcase #1\or)\or]\or\}\or\mathord>\or\rangle\or|\or\|\fi}
\nc\DElim[1]{\ifcase #1.\or(\or[\or\{\or<\or\langle\or|\or\|\fi}
\nc\delIM[1]{\ifcase #1.\or)\or]\or\}\or>\or\rangle\or|\or\|\fi}

\nc\cex{c_{\text{ex}}}
\nc\dd{\mathrm{d}}
\nc\etot{e_{\text{tot}}}  %% !!!! italic correction needed in these?
\nc\ekin{e_{\text{kin}}}
\nc\eint{e_{\text{int}}}
\nc\ethermal{e_{\text{thermal}}}
\nc\eel{e_{\text{el}}}
\nc\erheol{e_{\text{rheol}}}
\nc\je{j_e}
\nc\js{j_s}
\nc\jp{{j+1/2}}
\nc\jpp{{j+1}}
\nc\jm{{j-1/2}}
\nc\jmm{{j-1}}
\nc\np{{n+1/2}}
\nc\npp{{n+1}}
\nc\nm{{n-1/2}}
\nc\nmm{{n-1}}
\nc\myp{p}
\nc\myq{q}
\nc\Dt{\Delta t}
\nc\vb{v_{\text{b}}}
\nc\Dx{\Delta x}

\nc\Tex{T_{\text{ex}}}

\nc\pis{\pi_s}
\nc\sigmahat{\hat{\sigma}}
\nc\sigmael{\sigma_{\text{el}}}
\nc\qtaub{t_{\text{b}}}

\nc\Y{1. ex}  %% !!!!
\nc\Nl{\\[\Y]}
\nc\NnNl{\nonumber \Nl}
\nc\CoNl{, \NnNl}

\nc\VE{, \qquad}
\nc\VEE{, \nonumber \\[1. ex]}

\begin{document}

\title {Thermodynamically extended symplectic numerical simulation of viscoelastic, thermal expansion and heat conduction phenomena in solids}

 \author{Don\'at M. Tak\'acs, \'Aron Pozs\'ar, Tam\'as
F\"ul\"op\thanks{Corresponding author, \texttt{fulop.tamas@gpk.bme.hu}.}
 \\
 \\
Department of Energy Engineering, Faculty of Mechanical Engineering,
\\
Budapest University of Technology and Economics,
 \\
M\H uegyetem rkp.\ 3., H-1111 Budapest, Hungary;
 \\
Montavid Research Group, T\'ancos u.\ 6, H-1112, Budapest, Hungary
 }

\maketitle

\begin{abstract}
Symplectic numerical schemes for reversible dynamical systems predict the
solution reliably over large times as well, and are a good starting point for
extension to schemes for simulating irreversible situations like viscoelastic
wave propagation and heat conduction coupled via thermal expansion occuring
in rocks, plastics, biological samples etc.
Dissipation error (artificial nonpreservation of energies and amplitudes) of
the numerical solution should be as small as possible since it should not be
confused with the real dissipation occurring in the irreversible system. In
addition, the other well-known numerical artefact, dispersion error
(artificial oscillations emerging at sharp changes), should also be minimal
to avoid confusion with the true wavy behaviour.
The continuum thermodynamical aspects (respect for balances with fluxes,
systematic constitutive relationships between intensive quantities and
fluxes, the second law of thermodynamics with positive definite entropy
production, and the spacetime-based kinematic viewpoint) prove valuable for
obtaining such extended schemes and for monitoring the solutions.
Generalizing earlier works in this direction, here, we establish and
investigate such a numerical scheme for one-dimensional viscoelastic wave
propagation in the presence of heat conduction coupled via thermal expansion,
demonstrating long-term reliability and the applicability of
thermodynamics-based quantities in supervising the quality of the solution.
\end{abstract}


\section{Introduction}

Various kinds of solids, including rocks, plastics, and biological samples,
exhibit viscoelastic behaviour. This induces dissipation and an irreversible
source for temperature increase. Inhomogeneous temperature caused by this and
by external thermal effects---like thermal contact with a hydraulically
operated loading device, heat transfer and thermal radiation arriving from
the environment etc.---initiates heat conduction. Thermal expansion couples
heat conduction with the viscoelastic mechanics via the kinematic changes,
and it is this resulting coupled problem that one has to solve when wave
propagation and other time-dependent processes in such media are to be
described with sufficient accuracy. This `sufficient accuracy' may be a quite
demanding requirement, as for example in the problem of determining Newtonian
noise emerging from the gravitational effect of vibrations (mass
rearrangements) in the surrounding rock medium around underground
gravitational wave detectors like KAGRA and the planned third-generation
Einstein Telescope.

In order to obtain solutions, one way is to utilize available commercial
software.
Results of such studies can be found in \cite{fulop2020thermodynamical} and
\cite{pozsar2020four} where considerably simpler problems, elastic
propagation of a stress pulse in one and two spatial dimensions,
respectively, have been solved via numerous different built-in solution
methods of the finite-element software COMSOL. The results were
disappointing: some methods produced unstable outcome irrespective of
settings, some others
% exhibited
led to considerable dispersion error (artificial wavy patterns),
some
% presented
produced remarkable dissipation error (artificial damping), and the
more-or-less acceptable-looking solutions were considerably different so
there appeared a need for some tools to quantify the correctness of the
solution and thus to tell from the
% more-or-less acceptable-looking
solutions which may be closest to the correct one.
%but deeper and more
%quantitative analysis would require some appropriate additional tools.
 For a more complex problem with numerous fields and various couplings, this
path is more discouraging.

Another option is to develop an own numerical method, in a framework that
offers insight to the solution's quality in various ways, enabling
monitoring and controlling. It also should be reliable for large times (many
wave oscillations), artificial dissipation should be small as it should not
be confused with true---in our present case, viscoelastic and heat-conduction
related---dissipation, and dispersion error should also be small and
distinguishable from the true oscillations.

Choosing this latter option, in \cite{fulop2020thermodynamical},
% the first step has been taken, where,
 in one space dimension, large-time reliability of
elastic wave propagation was ensured by the symplectic Euler method,
reinterpreted as a space and time staggered finite-difference scheme, which
increased the first-order accuracy of the symplectic Euler method to
second-order. Notably, the reason why a symplectic method performs
outstandingly even over many time steps is that it proves to be the exact
integrator of a nearby other reversible (Hamiltonian) system (see,
\textit{e.g.,} \cite{hairer2006geometric}, p343).

\begin{figure}[t]
\centering
\parbox[c]{.25\columnwidth}{%
%\includegraphics[width=.24\columnwidth]{PTZ_circuit_1.pdf}%
%\vspace{5 ex}
\includegraphics[width=.24\columnwidth]{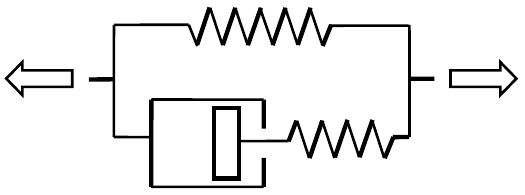}%
}
\hspace{.1\columnwidth}
\parbox[c]{.41\columnwidth}
{\includegraphics[width=.4\columnwidth]{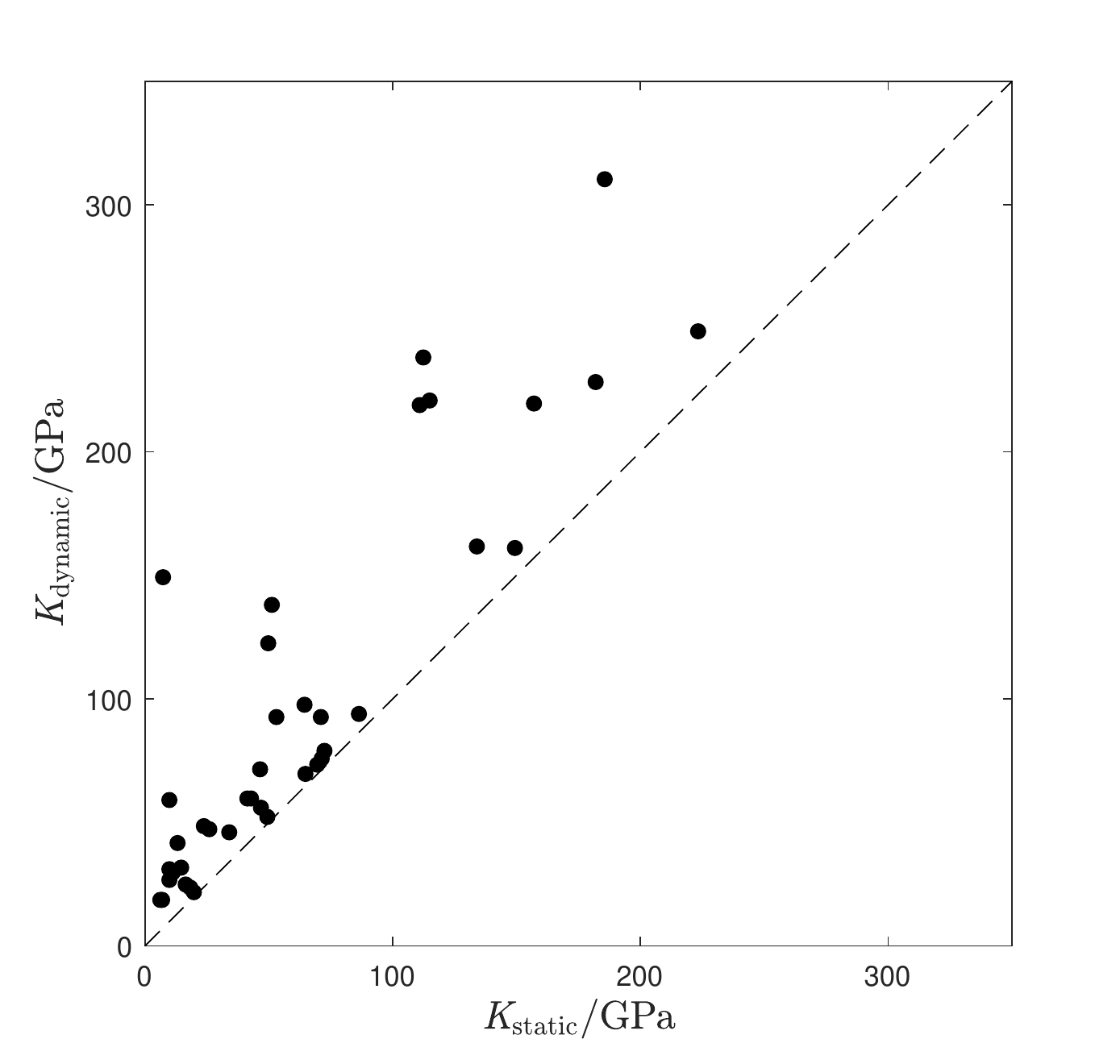}}
 \caption{%
 \textit{Left:}
%the two equivalent rheological circuit representations
 rheological circuit representation
of the classic (zero thermal expansion)
PTZ model \m { \sigma + \tau \pdv{\sigma}{t} = E \epsilon + \hat{E}
\pdv{\epsilon}{t} } between the stress-related quantity \m { \sigma } and the
strain-related \m { \epsilon }. This is the simplest model that exhibits both
creep and stress relaxation. The second law of thermodynamics requires \m {
\hat{I} \equiv \hat{E} - \tau E \ge 0 } or \m { \hat{E} \ge \tau E } among
the positive constant coefficients, inducing that the dynamic
(high-frequency) modulus \m { E_\infty = \hat{E} / \tau } is never smaller
than the static modulus \m {E }.
 \textit{Right:}
experimentally found relationship between dynamic and static bulk moduli for
various types of rock (data taken from \cite{davarpanah2020investigation}).
An analogous tendency is found for the shear moduli. Accordingly, many rocks
can successfully be modelled by a PTZ model in the volumetric/spherical part
and another PTZ model in the deviatoric part, in the spherical--deviatoric
decomposition \cite{asszonyi2015distinguished}.
 }  \label{PTZ}
 \end{figure}

 Also in
\cite{fulop2020thermodynamical}, this symplectic scheme
% has then been
 was
generalized for the Poynting--Thomson--Zener (PTZ) viscoelastic model (see
Figure~\ref{PTZ}). Stability analysis and investigation of the dispersion
relation have revealed the best settings
for having stability, no dispersion error, and minimal dissipation error. The
%generalized scheme was still second-order accurate both in space and in time.
generalization preserved second-order accuracy of the scheme both in space
and in time, and the finite-difference scheme remained explicit, making
simulations run fast, with low computational resource demand.

It is this numerical setup that has been used, in \cite{fulop2021wave}, for
numerically identifying the PTZ signal propagation velocity, finding good
agreement with the corresponding analytical results.

Next, in \cite{pozsar2020four}, the scheme was extended to three spatial
dimensions, for rectangular/cuboid geometries. The quantities provided by
thermodynamics were shown beneficial in monitoring the solution:
 \begin{itemize}
  \item
preservation of total energy demonstrated that the energy-friendly
 (Hamiltonian-friendly)
property
of the underlying symplectic scheme of the reversible part of the system was
preserved sucessfully during the generalization to the irreversible level;
  \item
the increase of temperature displayed where dissipation is the strongest; and
  \item
entropy production proved helpful in observing loss of stability early: its
turning into negative indicated violation of the second law of thermodynamics
and thus the emergence of instability.
 \end{itemize}

Here, we present the inclusion of thermal expansion and heat conduction into
the approach started in \cite{fulop2020thermodynamical}
and \cite{pozsar2020four}. The challenge is that temperature has to be raised to
the level of a fully active field that has its own dynamics (heat conduction)
and carries the coupling between this dynamics and the mechanical side via
its role in the kinematic aspects
through thermal expansion.

Can we still have an explicit finite-difference scheme, with second-order
accuracy in both space and time, with low dispersion error and dissipation
error? Additionally, can we further enhance the role of the
thermodynamics-provided quantities in monitoring the quality of the solution?

%Below,
 In what follows,
we find affirmative answers to these questions.

\section{%
%\hspace{-0.2em}
Continuum equations on the involved quantities} % !!!!

The thermodynamically consistent amalgamation of PTZ rheology, thermal
expansion, and heat conduction is the following set of continuum equations
(see detailed explanations below):
 \begin{align}  \label{@14038}
\varrho \dot{v} & = \pdv{\sigma}{x} ,
 \\  \label{@14321}
\sigma & = \sigmael + \sigmahat ,
 \\  \label{@15042}
L & = \pdv{v}{x} ,
 \\  \label{@15107}
\sigmahat + \tau \pdv{\sigmahat}{t} & = \hat{I} L ,
 \\  \label{@14341}
\etot & = \underbrace{\frac{v^2}{2}}_{\ekin}
 \mathrel  %% !!!!
+ \eint ,
 \\  \label{@14538}
\eint & = \underbrace{ \cex \left( T - \Tex \right) }_{\ethermal}
\mathrel  %% !!!!
+ \underbrace{ \frac{E}{2\varrho}
 \:\!  %% !!!!
\epsilon \left( \epsilon + 2 \alpha
\Tex \right) }_{\eel}
 \mathrel  %% !!!!
+ \underbrace{ \frac{\tau}{2 \varrho \hat{I}} \sigmahat^2 }_{\erheol}
 \\  \label{@15433}
s & = \cex \ln \frac{T}{\Tex} + \frac{E \alpha}{\varrho} \epsilon ,
 \\  \label{@15516}
\je & = - \lambda \pdv{T}{x} ,
 \\  \label{@15609}
\js & = \frac{1}{T} \je ,
 \\  \label{@15657}
\pis & = \pdv{\frac{1}{T}}{x} \je + \frac{1}{\hat{I}} \frac{1}{T}
 \;\!  %% !!!!
\sigmahat
\left( \hat{I} L - \tau \pdv{\sigmahat}{t} \right) ,
 \\  \label{@15924}
  \varrho \pdv{\eint}{t} &= - \pdv{\je}{x} + \sigma L ,
 \\  \label{@16027}
\varrho \pdv{s}{t} &= - \pdv{\js}{x} + \pis ,
 \\  \label{@16098}
\pdv{\epsilon}{t} & = L ,
 \\  \label{@16144}
\epsilon & = D + \alpha ( T - \Tex ) ,
 \\  \label{@16203}
\sigmael & = E D .
 \end{align}
Here, we work in one spatial dimension (\m { x }), in the small-strain regime
where, in leading order, density \m { \varrho } is constant and substantive
derivatives can be replaced by partial time derivatives \m { \pdv{}{t} }
(see, \textit{e.g.,} \cite{hetnarski2009thermal}). The balance of momentum
\re{@14038} determines the evolution of velocity \m { v } via the divergence
of stress \m {\sigma}. Stress is the sum of an elastic (reversible) term \m {
\sigmael } and an irreversible one \m { \sigmahat }, both explained below.
From velocity, the velocity gradient \m { L } is defined, with the aid of
which the Onsagerian time evolution of \m { \sigmahat } can be expressed as
\re{@15107} (\cite{asszonyi2015distinguished} applied to the PTZ case), in
which \m { \hat{I} } is positive (cf.\ the caption of Figure~\ref{PTZ}).

Furthermore, prior to the balance \re{@15924} of internal energy \m { \eint }
and the balance \re{@16027} of entropy \m { s }, the related constitutive
ingredients need to be provided: first, specific total energy \m { \etot } is
the sum of specific kinetic energy \m { \ekin } and specific internal energy
\m { \eint }. The form \re{@14538} of \m { \eint } originates from
\cite{lubarda2004thermodynamic}, then it is simplified along the lines of
\cite{nowacki1986thermoelasticity} (page 7), namely, we restrict ourselves to
a first-order expanded form in absolute temperature \m {T} around a chosen
temperature value \m { \Tex } (expansion point) with constant specific heat
capacity \m { \cex }.

Besides \m { T }, the second state variable of \m {
\eint } is the geometric quantity \m { \epsilon }, which equals strain when
the process starts from unstressed and thermodynamically equilibrial state at
\m { T = \Tex }. In \m{\epsilon}, we consider a second-order Taylor-expanded
form, the quadratic term providing the customary Hookean elastic energy
expression with Young's modulus \m{E}, while the term linear in \m{\epsilon}
gives rise to thermal expansion, with a constant thermal expansion
coefficient \m{\alpha}.

Finally, the third state variable is \m{\sigmahat}
itself (in a viscoelastic model beyond the level of PTZ, a more general
internal variable would stand here, see \cite{asszonyi2015distinguished},
Appendix~B), and the constant coefficient in \m { \erheol } is fixed by
thermodynamical consistency with \re{@15107} and \re{@15924}.
 Next, \m { s( T, \epsilon ) } is also fixed by thermodynamical consistency
(namely, the Gibbs relation with the reversible sum \m { \ethermal + \eel }).
Heat current density \m { \je } obeys Fourier's law \re{@15516} with heat
conduction coefficient \m { \lambda }, and entropy current density \m { \js }
is in the standard relationship \re{@15609} with \m { \je }. Entropy
production rate density \m { \pis } consists of two terms, the first proving
to be positive definite in light of \re{@15516} and the second because of
\re{@15107}, in accord with the second law of thermodynamics.

These have been the necessary ingredients for the balance of internal
energy \re{@15924} and for that of entropy \re{@16027}. Finally, the change
rate of \m { \epsilon } is
% expressed in terms of
 determined by
\m { L } and, as done in
\cite{hetnarski2009thermal} (cf.\ page 21), \m { \epsilon } is decomposed into
an elasticity-related part denoted here by \m { D } and a thermal part \m
{ \alpha ( T - \Tex ) }. \m { D } is the part that is related directly to
elastic stress, through \re{@16203}.
 We note that displacement \m { u } has not been used above but can be
calculated from velocity as \m { u_{t_0}^{} (t) = \int_{t_0}^{t} v \big(
 \!\;  %% !!!!
\tilde t
 \!\;  %% !!!!
\big)
 \!\:  %% !!!!
\dd \tilde t }.

This is the set of equations that is the generalization of the one treated in
\cite{fulop2020thermodynamical,fulop2021wave}, when thermal expansion and
heat conduction are added, both in their simplest form. Apparently, many
quantities play a role here. In such a case, a frequent approach is to
eliminate as many of the quantities as possible, reducing to a minimum the
number of degrees of freedom to be followed. Below, we prefer to keep all
these quantities as they provide flexibility in applying different boundary
conditions required by different applications, and as they are shown below to
prove useful for monitoring the quality of the solution. In case of such a
coupled problem, any insight of this kind is highly valuable.

\begin{table}
 \begin{center}
    \newcommand{\che}{{\color{gray}$D$, $\epsilon$, $\sigma$,
\color{gray}$\sigmahat$, $\sigmael$, $\eint$, $T$,} $L$}
    \newcommand{\cee}{$D$, $\epsilon$, $\sigma$, $\sigmahat$, $\sigmael$,
$\eint$, $T$, {\color{gray}$L$} }
    \newcommand{\chh}{$v$, {\color{gray}$\je$}}
    \newcommand{\ceh}{$\je$}

    \vspace*{1cm}
    \renewcommand{\arraystretch}{1.2}
    \begin{tabular}{
        >{\centering}m{0.7cm} |
        >{\centering}m{1.5cm}
        >{\centering}m{2.1cm}
        >{\centering}m{1.5cm}
        >{\centering}m{0.1cm}
        >{\centering}m{1.5cm}
        >{\centering}m{2.1cm}
        >{\centering}m{1.5cm}
        >{\centering}m{0.1cm}
    }
    \diagbox[height=8mm,width=11.2mm,innerleftsep=3.5mm]{$t$}{$x$} &
    $-\tfrac{1}{2}$ & 0 & +$\tfrac{1}{2}$ & $\cdots$ & $n-\tfrac{1}{2}$ & $n$
    & $n+\tfrac{1}{2}$ & $\cdots$ \tabularnewline[1.0ex] \hline
    $-\tfrac{1}{2}$ & \chh     & \che      & \chh     & $\cdots$ & \chh     &
    \che      & \chh     & $\cdots$ \tabularnewline[4ex]
    0               & \ceh     & \gc{\cee} & \ceh     & $\cdots$ & \ceh     &
    \gc{\cee} & \ceh     & $\cdots$ \tabularnewline[4ex]
    $+\tfrac{1}{2}$ & \chh     & \che      & \chh     & $\cdots$ & \chh     &
    \che      & \chh     & $\cdots$ \tabularnewline[1ex]
    $\vdots$        & $\vdots$ & $\vdots$  & $\vdots$ & $\ddots$ & $\vdots$ &
    $\vdots$  & $\vdots$ & $\cdots$ \tabularnewline[1ex]
    $j-\tfrac{1}{2}$& \chh     & \che      & \chh     & $\cdots$ & \chh     &
    \che      & \chh     & $\cdots$ \tabularnewline[4ex]
    $j$             & \ceh     & \gc{\cee} & \ceh     & $\cdots$ & \ceh     &
    \gc{\cee} & \ceh     & $\cdots$ \tabularnewline[4ex]
    $j+\tfrac{1}{2}$& \chh     & \che      & \chh     & $\cdots$ & \chh     &
    \che      & \chh     & $\cdots$ \tabularnewline[4ex]
    $j+1$           & \ceh     & \gc{\cee} & \ceh     & $\cdots$ & \ceh     &
    \gc{\cee} & \ceh     & $\cdots$ \tabularnewline[4ex]
    $\vdots$        & $\vdots$ & $\vdots$ & $\vdots$ & $\vdots$ & $\vdots$ &
    $\vdots$ & $\vdots$ & %$\ddots$
    \end{tabular}
 \end{center}
\caption{The finite-difference scheme. Grey quantities are intermediate ones
(not considered to be numerical solution values).
 }  \label{scheme}
\end{table}

\section{Discretization}

The space and time staggered finite-difference scheme applied here (see
Table~\ref{scheme}) is also the generalization of the one used in
\cite{fulop2020thermodynamical,fulop2021wave}. Accordingly, quantities reside
either at integer multiples \m { x_n = n \Delta x } of the space step \m {
\Delta x } or at half-integer ones \m { x_{n-1/2} = (n-1/2) \Delta x }.
Regarding time, not all quantities are classified as taken \textit{either} at
integer time steps \m { t^j = j \Delta t } \textit{or} at half-integer ones
\m { t^{j-1/2} = (j - 1/2) \Delta t }: some are calculated \textit{both} at
integer time steps \textit{and} at half-integer ones. In such cases, the ones
displayed in grey in Table~\ref{scheme} are not predicted values but
intermediate ones,
% somewhat like predictors in predictor--corrector schemes or, better, like
 similarly to
the
% intermediate quantities
 ones in Runge--Kutta-type schemes. They are introduced
because Fourier heat conduction breaks the leapfrog pattern that was present
in \cite{fulop2020thermodynamical,fulop2021wave} but we still aim at a
second-order accurate scheme.

Regarding initial conditions, we assume an equilibrial state (each of \m { v
}, \m { \sigmael }, \m { \sigmahat }, \m { L }, \m { \epsilon }, \m { D }, \m
{\je}, \m { \js }, \m { \pis } is zero) with homogeneous \m { T = \Tex } for
\m { t^0 } and \m { t^{-1/2} }, and, for definiteness, the boundaries \m {
x_{-1/2} }, \m { x_{N+1/2} } are thermally adiabatic and a single velocity
pulse 
 \begin{align}  \label{pulse}
v ( t, x_{-1/2} ) = \begin{cases}
\frac{\vb}{2} \left[ 1 - \cos \left( 2 \pi \frac{t}{\qtaub} \right) \right]
& \text{if} \quad 0 \le t \le \qtaub ,
 \\
0 & \text{otherwise}
\end{cases}
 \end{align}
is applied at the left endpoint, where \m { \qtaub } and \m { \vb } denote
the temporal width and the height of the pulse, respectively, while the right
endpoint is fixed (so \m { v = 0 }). Notably, these boundary conditions
forbid heat-type energy change and, after \m { \qtaub }, mechanical work is
also forbidden at both boundaries so total energy should be conserved after
the pulse, and even its increase during the pulse should be predicted as the
time integral of mechanical power at the left endpoint.

The explicit scheme conceived is the following chain of steps. Above each
equal sign, the corresponding continuum equation is displayed (if applicable:
for certain ``grey'' intermediate quantities there is no such equation).
As can be seen, the steps can be grouped into three blocks, the second and
third performing the same steps with
% a half time step shift.
 a shift of a half time step.
%Since each step is left--right balanced in space and earlier--later balanced
%in time, the scheme is second-order accurate both in space and time.
 \begin{gather}
v^\jp_\np \RR{@14038}  v^\jm_\np + \frac{1}{\varrho} \frac{\Dt}{\Dx}
\9 1 { \sigma^j_\npp - \sigma^j_n }
\VE
L^\jp_n \RR{@15042} \frac{1}{\Dx} \9 1 { v^\jp_\np - v^\jp_\nm }
\VE
\G{L}{j}{n} = \frac{1}{2} \9 1 { L^\jp_n + L^\jm_n }
 \VEE
\epsilon^\jpp_n \RR{@16098} \epsilon^j_n + \Dt L^\jp_n
 \VE
\G{\epsilon}{\jp}{n} = \frac{1}{2} \9 1 { \epsilon^\jpp_n + \epsilon^j_n }
 \VEE
\sigmahat^\jpp_n \RR{@15107} \frac{1}{ \frac{\tau}{\Dt} + \frac{1}{2} }
\9 2 { \hat{I} L^\jp_n + \9 1 { \frac{\tau}{\Dt} - \frac{1}{2} } \sigmahat^j_n }
 \VE
\G{\sigmahat}{\jp}{n} = \frac{1}{2} \9 1 { \sigmahat^\jpp_n + \sigmahat^j_n }
 , \label{@34846}
 \end{gather}
%qqqqwwww
 \begin{gather}
\G{\9 1 { \eint }}{\jp}{n} \RR{@15924} \G{\9 1 { \eint }}{\jm}{n} -
\frac{1}{\varrho} \frac{\Dt}{\Dx} \9 2 { \9 1 { \je }^j_\np - \9 1 { \je }^j_\nm } +
\frac{\Dt}{\varrho} \sigma^j_n \G{L}{j}{n}
 \VEE
\G{T}{\jp}{n} \RR{@14538} \Tex + \frac{1}{\cex}
\Bigg[
\G{\0 1 { \eint }}{\jp}{n}
- \frac{E}{2\varrho} \G{\epsilon}{\jp}{n} \0 1 { \G{\epsilon}{\jp}{n} +
2 \alpha \Tex
}
- \frac{\tau}{2 \varrho \hat{I}} \0 1 { \G{\sigmahat}{\jp}{n}
\rule{0pc}{3. ex}^2  %% !!!!
}
\Bigg]
 \VEE
\G{\9 1 { \je }}{\jp}{\np} \RR{@15516} - \frac{\lambda}{\Dx}
\9 1 { \G{T}{\jp}{\npp} - \G{T}{\jp}{n} }
 \VEE
\G{D}{\jp}{n} \RR{@16144} \G{\epsilon}{\jp}{n} -
\alpha \9 1 { \G{T}{\jp}{n} - \Tex }
 \VE
\G{\9 1 { \sigmael }}{\jp}{n} \RR{@16203} E \G{D}{\jp}{n}
 \VE
\G{\sigma}{\jp}{n} \RR{@14321} \G{\9 1 { \sigmael }}{\jp}{n} +
\G{\sigmahat}{\jp}{n}
 ,
 \label{@27250}
% \CoNl
 \end{gather}
%qqqqwwww
 \begin{gather}
\9 1 { \eint }^\jpp_n \RR{@15924} \9 1 { \eint }^j_n - \frac{1}{\varrho}
\frac{\Dt}{\Dx} \left\{ \9 2 { 2 \9 1 { \je }^j_\np - \G{\9 1 { \je }}{\jm}{\np}
} \right.
- \left. \9 2 { 2 \9 1 { \je }^j_\nm - \G{\9 1 { \je }}{\jm}{\nm} } \right\} +
\frac{\Dt}{\varrho} \G{\sigma}{\jp}{n} L^\jp_n
 \VEE
T^\jpp_n \RR{@14538} \Tex + \frac{1}{\cex} \9 2 {
 \9 1 { \eint }^\jpp_n - \frac{E}{2\varrho} \epsilon^\jpp_n
 \9 1 { \epsilon^\jpp_n + 2 \alpha \Tex } - \frac{\tau}{2 \varrho \hat{I}}
 \9 1 { \sigmahat^\jpp_n }^2
}
 \VEE
\9 1 { \je }^\jpp_\np \RR{@15516} - \frac{\lambda}{\Dx} \9 1 { T^\jpp_\npp -
T^\jpp_n }
 \VEE
D^\jpp_n \RR{@16144} \epsilon^\jpp_n - \alpha \9 1 { T^\jpp_n - \Tex }
 \VE
\9 1 { \sigmael }^\jpp_n \RR{@16203} E D^\jpp_n
 \VE
\sigma^\jpp_n \RR{@14321} \9 1 { \sigmael }^\jpp_n + \sigmahat^\jpp_n
 .
 \label{@37685}
 \end{gather}

%Since each step is left--right balanced in space and earlier--later balanced
%in time, the scheme is second-order accurate both in space and time.
This scheme is second-order accurate both in space and time. This can be seen
as follows. Each of the formulae in \re{@34846}--\re{@37685} is devised in
such a way that it has a time instant and a space location around which
each of the terms in that formula provides an accurate approximation of the
corresponding continuum term not only at leading-order but also at
next-to-leading order, when expanding into Taylor series in time and space
both. For example, the third equation in the group \re{@34846} is centered
around \m { t^j } and \m { x_n }, and the expansion of the average on the
right-hand side gives
 \begin{align}
& \frac{1}{2} \9 1 { L^\jp_n + L^\jm_n } = \frac{1}{2} \9 2 {
L \0 1 { t^\jp, x_n } + L \0 1 { t^\jm, x_n } }
 \nonumber \\
& = \frac{1}{2} \9 3 { L \0 1 { t^j + \frac{\Delta t}{2}, x_n } +
L \0 1 { t^j - \frac{\Delta t}{2}, x_n } }
 \nonumber \\
& = \frac{1}{2} \left\{ \9 2 { L \0 1 { t^j, x_n } + \pdv{L}{x}
\0 1 { t^j, x_n } \frac{\Delta t}{2} + \mathcal{O} \0 1 { \Delta t^2 } } \right.
 \nonumber \\
& \quad + \left. \9 2 { L \0 1 { t^j, x_n } - \pdv{L}{x}
\0 1 { t^j, x_n } \frac{\Delta t}{2} + \mathcal{O} \0 1 { \Delta t^2 } } \right\}
 \nonumber \\
& = L \0 1 { t^j, x_n } + \mathcal{O} \0 1 { \Delta t^2 }
= \G{L}{j}{n} + \mathcal{O} \0 1 { \Delta t^2 } .
  \label{@37970}
 \end{align}

Next, average is actually a special case of linear interpolation and a more
general statement also holds: any linear interpolation or linear
extrapolation is accurate not only at leading order but also at
next-to-leading order: \textit{e.g.,} the interpolation/extrapolation of an
\m { f \1 1 {t} } from \m { f\1 1 {t - \myp \Delta t} } and \m { f\1 1 {t -
\myq \Delta t} } \1 1 {here, \m { p, q } may be negative} is
 \begin{align}
& f \1 1 {t - \myp \Delta t} + \frac{ f \1 1 {t - \myp \Delta t} -
f \1 1 {t - \myq \Delta t} }{ \1 1 { t - \myp \Delta t } - \1 1 {t -
\myq \Delta t} } \9 2 {
 \rule{0pc}{2 ex}  %% !!!!
t - \9 1 { t - \myp \Delta t } }
 \nonumber \\
 \label{@38847}
 &
 \qquad
%= \cdots
= \frac{\myq}{\myq - \myp} f \1 1 {t - \myp \Delta t}
- \frac{\myp}{\myq - \myp} f \1 1 {t - \myq \Delta t}
= \cdots = f \1 1 {t} + \mathcal{O} \0 1 { \Delta t^2 } .
 \end{align}
Such extrapolations are applied in the first equation in the group
\re{@37685}, extrapolating \m { \9 1 { \je }^\jp_\np } from \m { \9 1 { \je
}^j_\np } and \m { \9 1 { \je }^\jm_\np }, and extrapolating \m { \9 1 { \je
}^\jp_\nm } from \m { \9 1 { \je }^j_\nm } and \m { \9 1 { \je }^\jm_\nm } \1
1 {case \m { p = 1/2 }, \m { q = 1 }}.

Further, the
 difference-ratio
approximation of a space derivative is also accurate even at next-to-leading
order around the midpoint: \textit{e.g.,}
 \begin{align}  \label{@39826}
\frac{\sigma^j_\npp - \sigma^j_n}{\Dx} = \cdots = \pdv{\sigma}{x}
\0 1 { t^j, x_\np } + \mathcal{O} \0 1 { \Delta x^2 } .
 \end{align}
The analogous statement holds for time derivatives.
 \OMIT{%
Then, when a time
derivative containing formula is rewritten like
 \begin{align}  \label{@40142}
\frac{\epsilon^\jpp_n - \epsilon^j_n}{\Delta t} = L^\jp_n
 \quad \Longrightarrow \quad
\epsilon^\jpp_n = \epsilon^j_n + \Dt L^\jp_n ,
 \end{align}
%the resulting formula has
the error
% changes as \m { \mathcal{O} \0 1 { \Delta t^2 } \rightarrow
of the new formula is \m {
\mathcal{O} \0 1 { \Delta t^3 } } so, during one time step, the approximation
is accurate up to second order, \textit{i.e.,} up to \m { \Delta t^2 }.
 }%(V)OMIT
 \VOMIT{%
Then, if a formula approximating a derivative with an error of  \m {
\mathcal{O} \0 1 { \Delta t^2 } } is rewritten to give the time evolution of
the field, the error of this approximation is \m {
\mathcal{O} \0 1 { \Delta t^3 } }. For example,
 \begin{align}  \label{@40142}
\frac{\epsilon^\jpp_n - \epsilon^j_n}{\Delta t} = L^\jp_n
 \quad \Longrightarrow \quad
\epsilon^\jpp_n = \epsilon^j_n + \Dt L^\jp_n
 \end{align}
%gives
 yields
an approximation for the
% new
 change of the
\m { \epsilon } value that is accurate up
to second order, \textit{i.e.,} up to \m { \Delta t^2 }.
 }%(V)OMIT

This conclusion can also be demonstrated numerically, as we can see in the
subsequent Section.

\section{Simulation results}

In the calculations, quantities have been made dimensionless (in notation:
overtilde) with respect to the sample length \m { X } as length unit, the
travel time along this length of the rheological wave with the maximal speed
\m { \hat{c} = \sqrt{ \frac{ \hat{E}/\tau }{\varrho} } } as the time unit, \m
{ E } \1 1 {which fixes the mass unit}, and \m { \cex } \1 1 {which fixes the
%
%temperature unit}. Unless stated otherwise, the rheological Courant number \m
%% TD kalapálása, FT átkalapálásában:
temperature unit}. Correspondingly, the dimensionless parameters and
variables are
 \begin{gather}
    \tilde{t} = \frac{\hat{c}}{X} t , \quad
    \Delta \tilde{t} = \frac{\hat{c}}{X} \Delta t , \quad
    \tilde{\tau} = \frac{\hat{c}}{X} \tau , \quad
    \tilde{t}_{\text{b}} = \frac{\hat{c}}{X} \qtaub ,
% \quad
 \\
    \tilde{x} = \frac{1}{X} x , \quad
    \Delta \tilde{x} = \frac{1}{X} \Delta x , \quad
    \tilde{X} = \frac{X}{X} = 1 , \quad
    \tilde{\hat{c}} = \frac{\hat{c}}{\hat{c}} = 1 , \quad
    \tilde{v}_{\text{b}} = \frac{1}{\hat{c}} \vb , \quad
 \\
    \tilde{\varrho} = \frac{\hat{c}^2}{E} \varrho , \quad
    \tilde{E} = \frac{E}{E} = 1 , \quad
    \tilde{\hat{E}} = \frac{\hat{c}}{X E} \hat{E} , \quad
    \tilde{\hat{I}} = \frac{\hat{c}}{X E} \hat{I} , \quad
 \\
    \tilde{\lambda} = \frac{\hat{c}}{E X \cex} \lambda , \quad
    \tilde{T}_{\mathrm{ex}} = \frac{\cex}{\hat{c}^2} \Tex , \quad
    \tilde{c}_{\mathrm{ex}} = \frac{\cex}{\cex} = 1 , \quad
    \tilde{\alpha} = \frac{\hat{c}^2}{\cex} \alpha .
 \end{gather}
Further, the nondimensional counterparts of the various field variables are
 \begin{gather}
    \tilde{D} = D , \quad
    \tilde{\epsilon} = \epsilon , \quad
    \tilde{v} = \frac{1}{\hat{c}} v , \quad
    \tilde{L} = \frac{X}{\hat{c}} L , \quad
    \tilde{\sigma}_{\text{el}} = \frac{1}{E} \sigmael , \quad
    \tilde{T} = \frac{\cex}{\hat{c}^2} T , \quad
 \\
    \tilde{e}_{\mathrm{int}} = \frac{1}{\hat{c}^2} \eint , \quad
    \tilde{\je} = \frac{1}{E \hat{c}} \je , \quad
    \tilde{s} = \frac{1}{\cex} s , \quad
    \tilde{j}_{s} = \frac{\hat{c}}{E \cex} \js, \quad
    \tilde{\pi}_{s} = \frac{X \hat{c}}{E \cex} \pis
 \end{gather}
(and, for the other specific energy and stress quantities, accordingly). The
dimensionless versions of \re{@14038}--\re{@16203} and
\re{@34846}--\re{@37685} are the straightforward counterparts where a tilde
is applied above all quantities.

%Unless stated otherwise, the rheological Courant number \m
%% end of TD kalapálása

During testing, we have scanned a large part of the space of nondimensional
parameters and report here all the observed features we have found important
to mention.
 Unless stated otherwise, the rheological Courant number \m
{ \hat{C} = \hat{c} \frac{\Dt}{\Dx} } has been set to \m { 1 }, \m { N = 30 }
space cells have been used, the simulated dimensionless time was \m { 15
}, and $\tilde{v}_{\mathrm{b}}=0.01$, $\tilde{t}_{\mathrm{b}}=0.5$,
$\tilde{T}_{\mathrm{ex}}=1$ have been used.
 \1 1 {Details on why \m { \hat{C} = 1 } appears the best choice follow
later.}
 Five materials \1 1 {Stripa granite
\cite{swan1978mechnical,chan1982rock,rockstudy2022rock}, plastic PA6
\cite{goodfellow2022polyamide,asszonyi2016elastic}, and three artificial ones
for probing certain extremities} have been considered, deliberately chosen
%to see the effects of rather different material properties
to examine various circumstances induced by the rather different material
properties.
% Note that it is not easy to find real materials for which all of the needed
%coefficients are available from the literature.
% Nevertheless, with
 With the inclusion of the three artificial materials, we can
demonstrate all the remarkable effects we have found.
 \OMIT{%
We present results for three materials \1 1 {Stripa granite
\cite{swan1978mechnical,chan1982rock,rockstudy2022rock}, plastic PA6
\cite{goodfellow2022polyamide,asszonyi2016elastic}, and an artificial one for
probing certain extremities}%
% have been considered
, deliberately chosen
%to see the effects of rather different material properties
to test various circumstances induced by the rather different material
properties.
 Note that it is not easy to find real materials for which all of the needed
coefficients are available from the literature. Nevertheless, we have
actually scanned a large part of the space of nondimensional parameters and
report here all the observed features we have found important to mention.
 }%(V)OMIT

\begin{table}[b]
    \centering
    \begin{tabular}{c c c c c c c c c}
        \toprule
          & $E$ &  $\hat{I}$ & $\alpha$ & $\cex$ & $\lambda$ 
          % & $\nu$ 
            & $\rho$ & $\tau$ \\
        \midrule
          & \si{\pascal} & \si{\pascal\second} & \si{1\per{\kelvin}} &\si{\joule\per{(\kilogram\kelvin)}} & \si{\watt\per{(\meter\kelvin)}} 
          % & \si{1} 
          & \si{\kilogram\per\meter^{3}} & \si{\second} \\
        \midrule
%        A & \num{0.5} & \num{+6.2500e-01} & \num{+1.0000e-03} & \num{1} & \num{0.01} 
%          % & \num{0.3} 
%          & \num{1} & \num{+1.2500e+00} \\
%        B & \num{0.5} & \num{+6.2500e-01} & \num{+1.0000e-03} & \num{1} & \num{0.03} 
%          % & \num{0.3} 
%            & \num{1} & \num{+1.2500e+00} \\
        PA6 & \num{+1.1806e+09} & \num{+2.2103e+08} & \num{+8.0000e-07} & \num{+1700.0} & \num{0.26} 
            % & \num{0.37086}
            & \num{+1150.0} & \num{0.76750} \\
        SG & \num{+5.1300e+10} & \num{+9.0863e+13} & \num{+1.1100e-05} & \num{+804.29} & \num{3.1} 
           % & \num{0.23000} 
           & \num{+2622.0} & \num{1108.8} \\
        \bottomrule \\
    \end{tabular}
    \caption{%
%Material
Properties of the real materials used for the various simulations.}\label{tab:material_parameters}
\end{table}

\begin{table}[b]
    \centering
    \begin{tabular}{c c c c c c c c c}
        \toprule
          & $\tilde{E}$ &  $\tilde{\hat{I}}$ & $\tilde{\alpha}$ & $\tilde{c}_{\mathrm{ex}}$ & $\tilde{\lambda}$ 
          % & $\tilde{\nu}$ 
          & $\tilde{\rho}$ & $\tilde{\tau}$ \\
        \midrule
        A & \num{+1.0000e+00} & \num{+1.2500e+00} & \num{+1.0000e-03} & \num{+1.0000e+00} &
        \num{+4.0000e-04}
          % & \num{0.3} 
        & \num{+2.0000e+00} & \num{+1.2500e+00} \\
        B & \num{+1.0000e+00} & \num{+1.2500e+00} & \num{+1.0000e-03} & \num{+1.0000e+00} & \num{+2.0000e-02} 
          % & \num{0.3} 
        & \num{+2.0000e+00} & \num{+1.2500e+00} \\
        C & \num{+1.0000e+00} & \num{+1.2500e+00} & \num{+1.0000e-03} & \num{+1.0000e+00} & \num{+6.0000e-02} 
          % & \num{0.3} 
        & \num{+2.0000e+00} & \num{+1.2500e+00} \\
        PA6 & \num{+1.0000e+00} & \num{+2.1157e+03} & \num{+6.0095e-04} & \num{+1.0000e+00} & \num{+1.4640e-09} 
          % & \num{0.37086} 
        & \num{+1.2439e+00} & \num{+8.6732e+03} \\
        SG & \num{+1.0000e+00} & \num{+1.2626e+07} & \num{+7.0135e-01} & \num{+1.0000e+00} & \num{+5.3561e-10} 
           % & \num{0.23000} 
         & \num{+2.5974e+00} & \num{+7.9043e+06} \\
        \bottomrule \\
    \end{tabular}
   \caption{Nondimensional material properties used for the various
 simulations.}\label{tab:nondim_material_parameters}
\end{table}

\OMIT{
\begin{table}
    \centering
    \begin{tabular}{c c c c c c c c}
        \toprule
        & $X$ (\si{\meter}) & $\Tex$ (\si{\kelvin}) & $\tilde{t}_b$ & $\tilde{t}_{\mathrm{end}}$ & $\tilde{v}_{\mathrm{max}}$ & $\hat{C}$ & $N$\\
        \midrule 
        Fig.~\ref{fast_diss_good_heat_cond_v_T} & 1 & 1 & 0.5 & 15 & 0.01 & 1.0 & 30 \\
        Fig.~\ref{fast_diss_no_heat_cond_v_T} & 1 & 1 & 0.5 & 15 & 0.01 & 1.0 & 30 \\
        Fig.~\ref{pa6_dispersion} & 0.1 & 293 & 0.5 & 15 & 0.01 & 0.5 & 30 \\
        Fig.~\ref{unstable_lambda} & 1 & 1 & 0.5 & 5 & 0.01 & 1 & 30 \\
        \bottomrule \\
    \end{tabular}
    \caption{Summary of simulation parameters.}
 \label{tab:simulation_parameters}
\end{table}
}

\begin{figure}[H]
     \centering
     \includegraphics{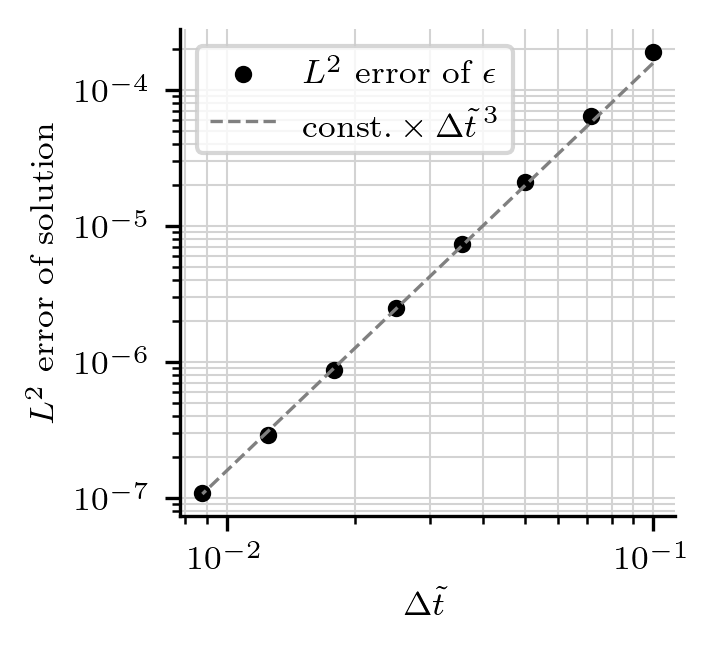}%
     \includegraphics{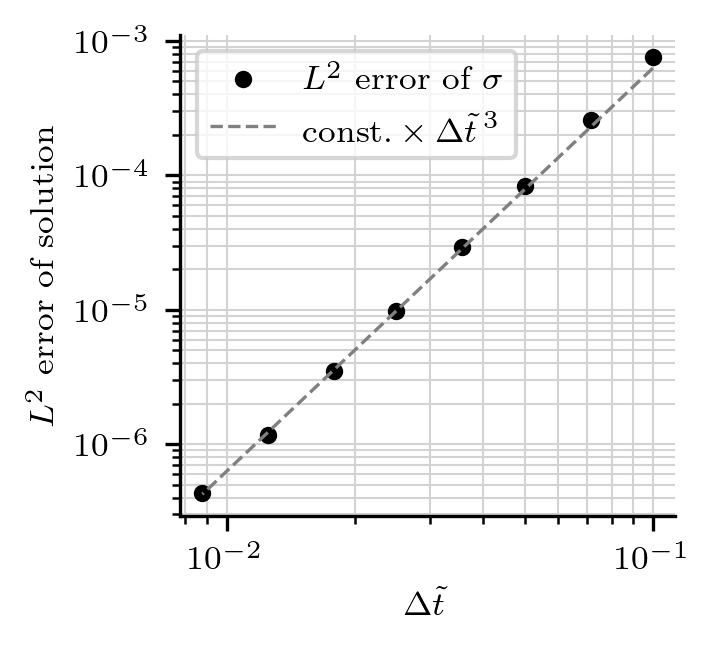}
     \caption{Numerical evidence demonstrating second-order accuracy  
of the numerical scheme. Displayed is the \m { L^2 } error along the sample
during one time step, shown for two quantities \1 1 {presentation inspired by
\cite{love2013convergence}}. The error appears at the order of \m { \Delta
t^3 }. Calculation done with material A \1 1 {see
Table~\ref{tab:nondim_material_parameters}}.}
 \label{convergence}
 \end{figure}

First presented is the numerical demonstration of the second-order accuracy
of the numerical scheme. As visible in Figure~\ref{convergence}, the
simulation result agrees with the analytical one of the previous Section.

\begin{figure}[H]
 \centering
 \includegraphics[height=25ex]{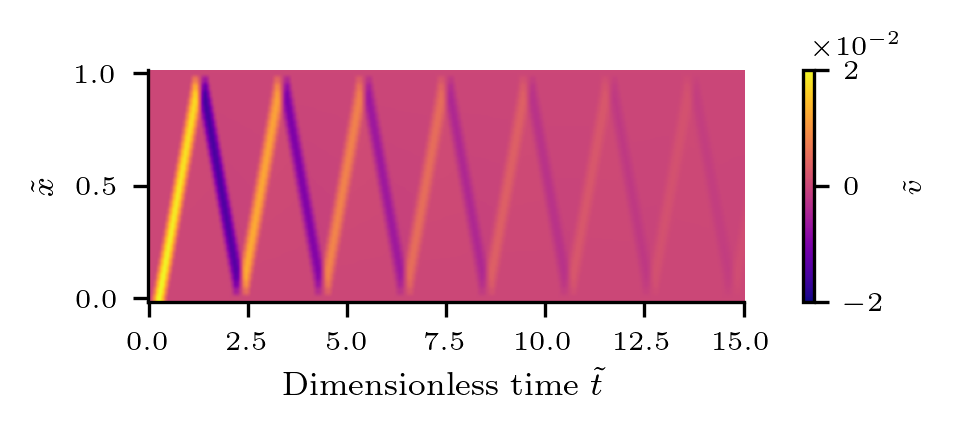}
 \hfill
 \includegraphics[height=25ex]{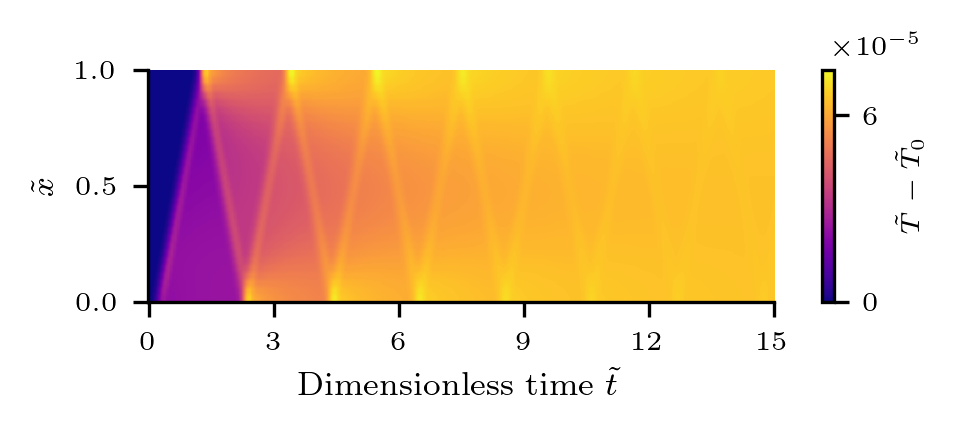}
 \\
 \hspace{6. em}\includegraphics[height=42ex]{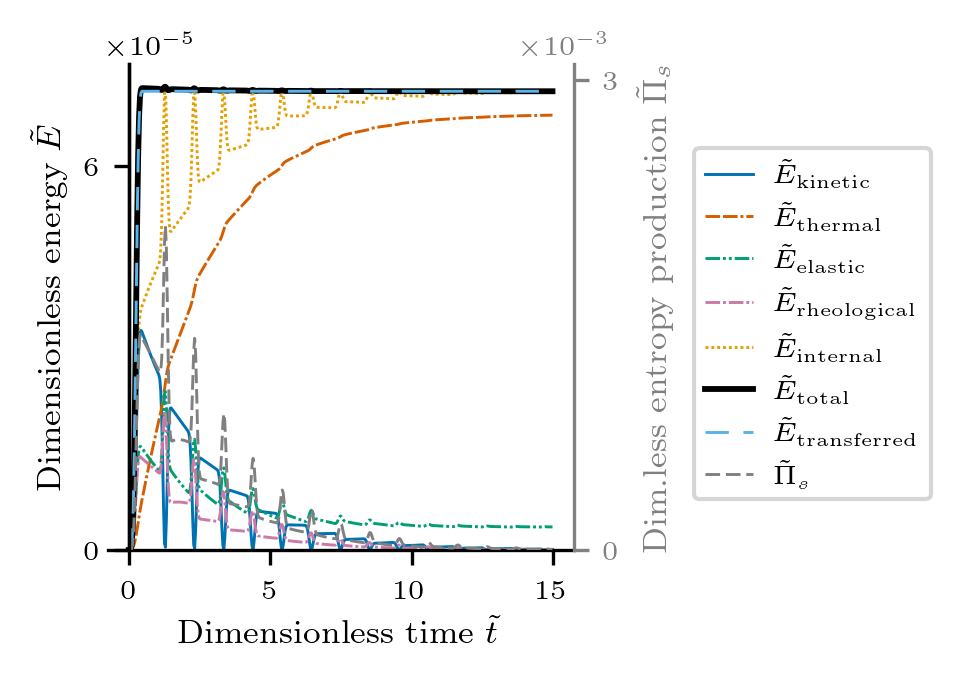}
 \caption{
 \textit{First row:}
Space-time dependence of velocity (left) and of temperature (right) when both
mechanical dissipation and heat conduction are strong. Other mechanical and
kinematic quantities behave similarly to velocity. Temperature inhomogeneity
created by dissipation (which is concentrated along the path of the
travelling pulse) gets homogenized by heat conduction.
 \textit{Second row:}
time dependence of the corresponding various energies (space integrals) and
of entropy production rate (space integral of \m { \pis }).
 Results shown for
%Calculation done with
 material B \1 1 {see Table~\ref{tab:nondim_material_parameters}}. }
  \label{fast_diss_good_heat_cond_v_T}
 \end{figure}

In the next group of results shown \1 1 {see
Figure~\ref{fast_diss_good_heat_cond_v_T}}, artificial material B has been
chosen, in order to make the effects of the coupled phenomena fully visible.

As can be seen, total energy is preserved reliably, and even its initial
increase during the pulse agrees with the transferred energy (time integrated
mechanical power at the boundary). All energies tend to their expected
equilibrial asymptotic value. As a special case, elastic energy tends to a
nonzero constant since dissipation raised temperature, which generated a
nonzero \m { \epsilon } via thermal expansion. In parallel, entropy
production rate is reassuringly positive, in consistency with the continuum
expectation due to the second law of thermodynamics.

\begin{figure}[H]
 \centering
 \includegraphics[height=25ex]{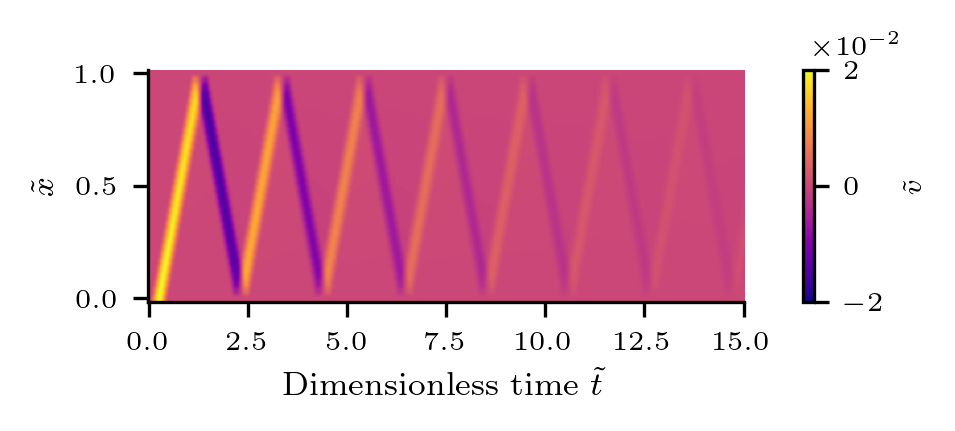}
 \hfill
 \includegraphics[height=25ex]{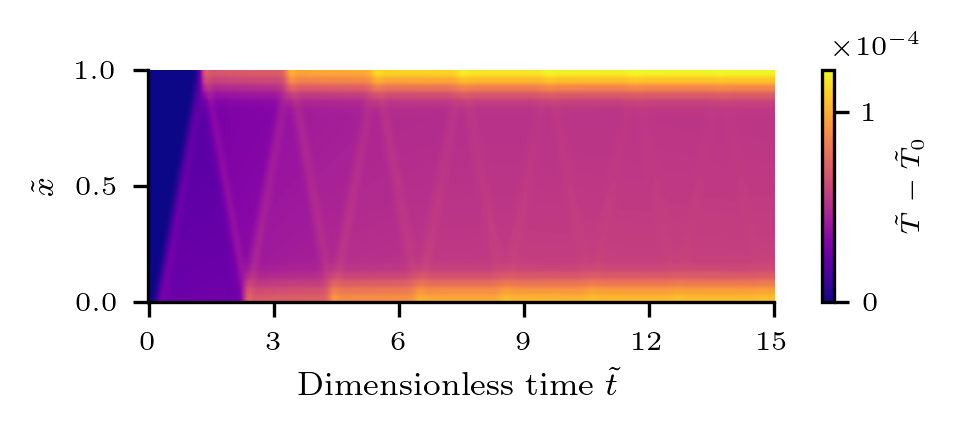}
 \\
 \hspace{6. em}\includegraphics[height=42ex]{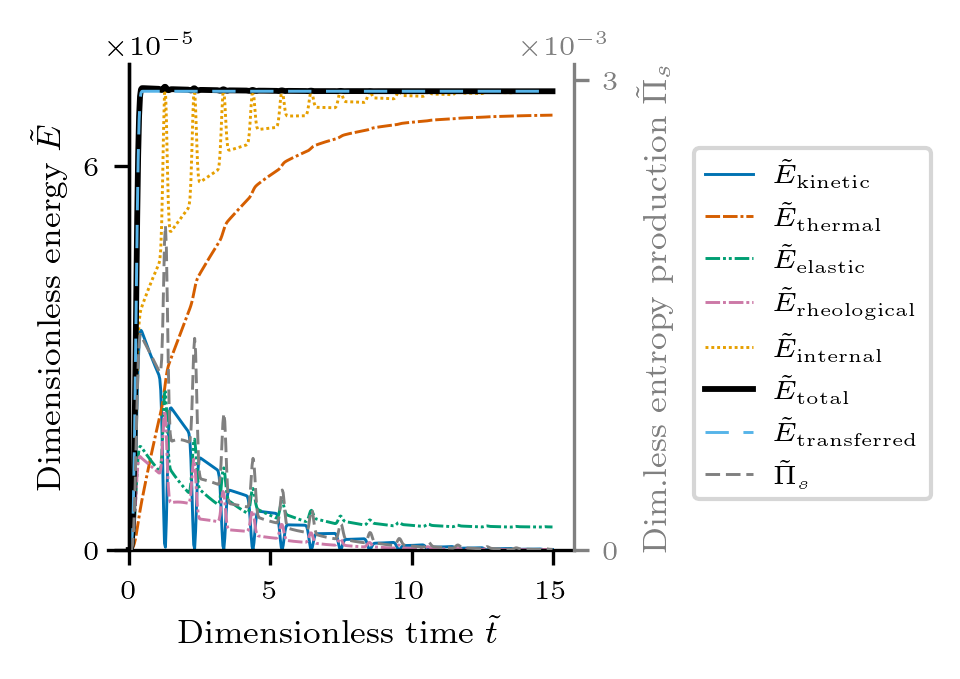}
 \caption{
%\textbf{B anyag}
Results analogous to those in Figure~\ref{fast_diss_good_heat_cond_v_T} but
with no heat conduction.
 }  \label{fast_diss_no_heat_cond_v_T}
 \end{figure}

For comparison, these calculations have been repeated with heat conduction
``switched off''; see Figure~\ref{fast_diss_no_heat_cond_v_T} for the
differences.
The only obvious difference is that the distribution of temperature cannot
get homogenized. The turning points create the most dominant amount of
dissipation (viscoelasticity is most active where changes are the most
considerable) and temperature increase remains localized there. In contrast,
within the bulk, the many back-and-forth travels of the pulse generate
dissipation approximately homogeneously.

For the Stripa granite and for the PA6 plastic, heat conduction was found to
produce a weak effect, somewhat similarly to the outcomes shown in
Figure~\ref{fast_diss_no_heat_cond_v_T}. The other possible source for
differences was the interplay among the viscoelastic creep and relaxation
time scales, the travel time of the pulse along the sample, and the duration
of the pulse. In many realistic cases, dissipation became visible only after
many bounces. Naturally, the effect of viscoelasticity is visible even in
such cases as the propagation speed of the pulse, \m { \hat{c} = \sqrt{ \frac{
\hat{E}/\tau }{\varrho} } }, is larger than the elastic/low-frequency value
\m { c = \sqrt{ \frac{E}{\varrho} } }.

\begin{figure}[H]
 \centering
 \includegraphics[height=25ex]{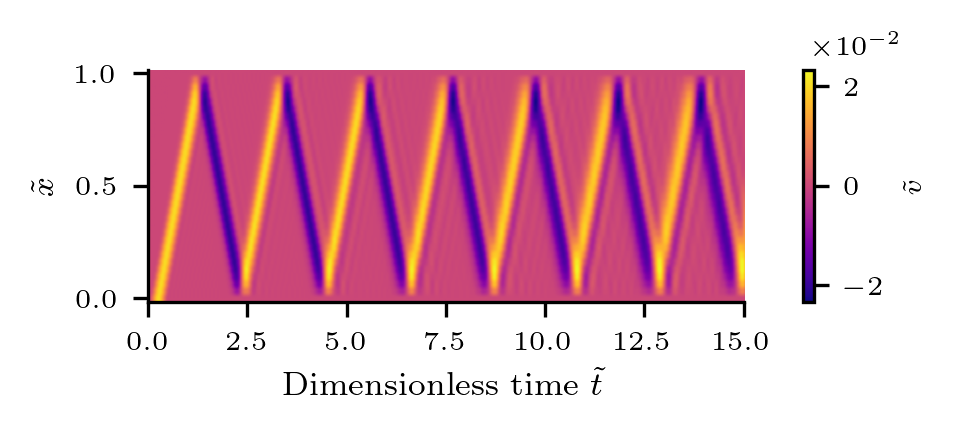}
 \hfill
 \includegraphics[height=25ex]{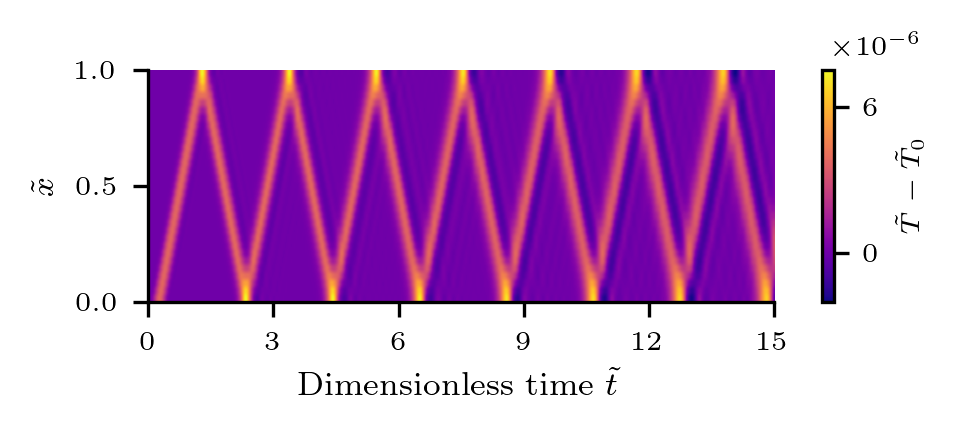}
 \\
 \hspace{6. em}\includegraphics[height=42ex]{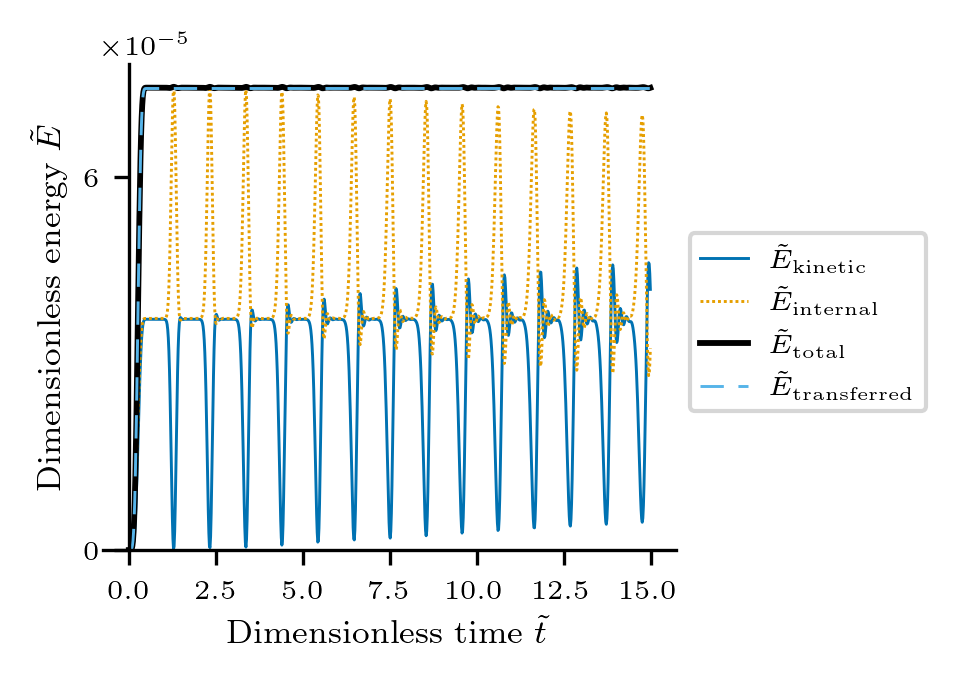}
 \caption{
With Courant number \m { \hat{C} = 0.5 }, dispersion error is observable.
Note the gradually emerging ghost images around the pulse in the space-time
diagrams, and the artificial oscillations in internal and kinetic energies.
Result shown for material PA6
\1 1 {see Table~\ref{tab:nondim_material_parameters}}.
 }  \label{pa6_dispersion}
 \end{figure}

%Presented
% Follows next is
Next follows
 the study of dispersion error.
 In
Figure~\ref{pa6_dispersion}, outcomes are shown for PA6 plastic, with the
rheological Courant number lowered from \m { 1 } to \m { 0.5 }, at room
temperature ($\tilde{T}_{\mathrm{ex}}=0.39$).
%, with a realistic sample size of $X=\SI{0.1}{\meter}$.
No problem is
visible in the total energy, however, the various energy terms unveil the
presence of dispersion error. Moreover, the dubious change in the
maxima/minima of the energy terms raises concern regarding dissipation error.

%Experience with such examples it appears to indicate
 This and similar results
% appear to
 visually indicate
that, for this solid mechanical model, rheological Courant number~\m { 1 } is
a superior choice compared to other choices.
 Actually, one can also have a more quantitative support for this
%conjecture.
 conviction.
Namely, a starting point is the previous works \cite{fulop2020thermodynamical,
pozsar2020four} \1 1 {studying the case of no heat conduction and no thermal
expansion} where it is not only observed but also verified, via analysing the
dispersion relation, that Courant number 1 is the optimal
value regarding dispersion error. Due to the
parabolic nature of heat conduction, a plausible expectation is that the same
may apply here as well.
Here, analytical studies would be much more involved
% since the set of equations is nonlinear.
 because of the more complicated model so we include a numerical study where
the Courant number is scanned between \m { 0.3 } and \m { 1 } while all other
parameters \1 1 {including the space step} are kept fixed.
\1 1 {The time step is changed according to \m { \Dt = \frac{\hat{C}}{\hat{c}} \Dx
}.}

 \begin{figure}[H]
     \centering
     \includegraphics{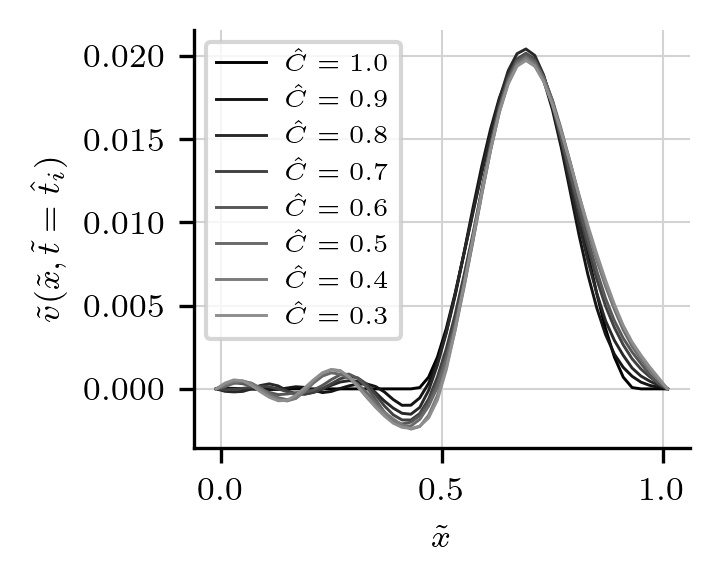}%
     \includegraphics{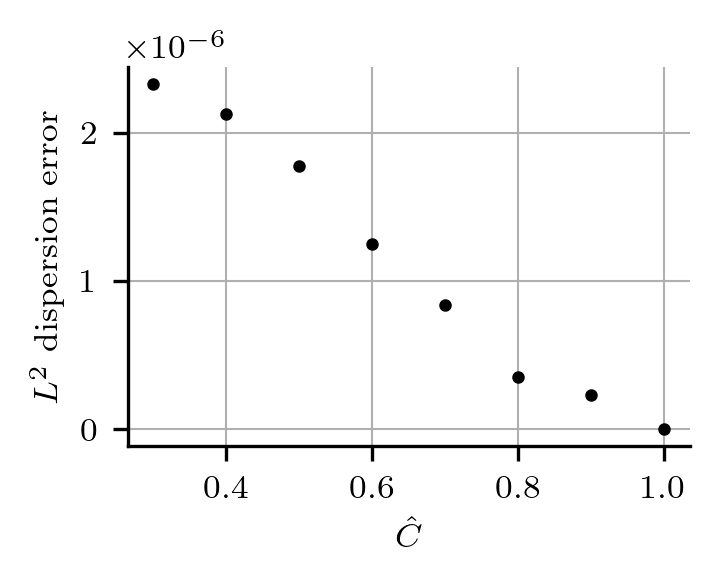}
     \caption{%
\textit{Left:} Courant number dependence of the shape of the propagating
excitation. As \m { \hat{C} } is gradually decreased from 1, the artificial
oscillation
% around
 behind the pulse gets larger and larger.
  \textit{Right:} the \m { L^2 }
% measure of dispersion error
deviation of the
% pattern
excitation shape
% seen on the left
 from that of the \m { \hat{C} = 1 } one.
Calculation performed with material PA6%
%\ \1 1 {see Table~\ref{tab:nondim_material_parameters}}%
.}
 \label{dispersion_courant}
 \end{figure}

Now, looking back to Figure~\ref{dispersion_courant}, we can see that, with
optimal numerical settings, the scheme is energy-friendly not only in the
sense of total energy conservation but the various energy terms themselves
are also reliably calculated; while for suboptimal numerical settings the
various energy terms display the appearance of numerical artefacts.

Finally, instability is investigated. In our coupled problem, it is not only
the Courant number that poses a constraint on the time and space steps but
heat conduction also brings in its stability criterion. 

 \begin{figure}[t]
 \centering
 \includegraphics[height=25ex]{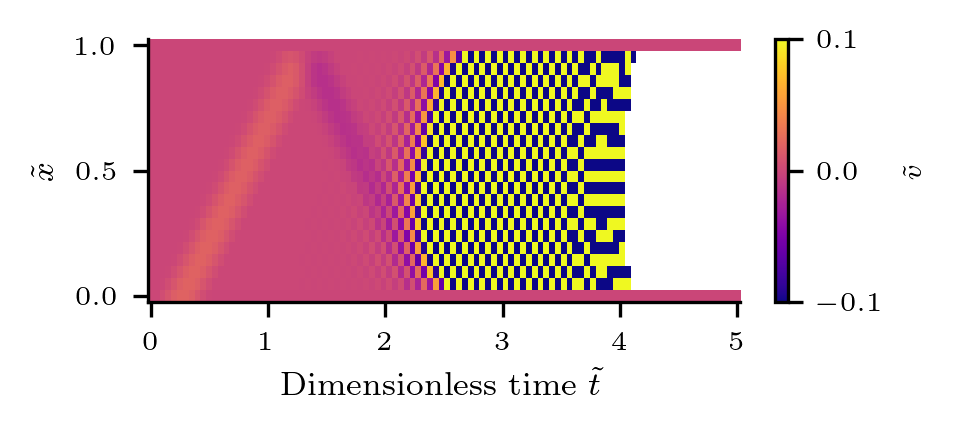}
 \hfill
 \includegraphics[height=25ex]{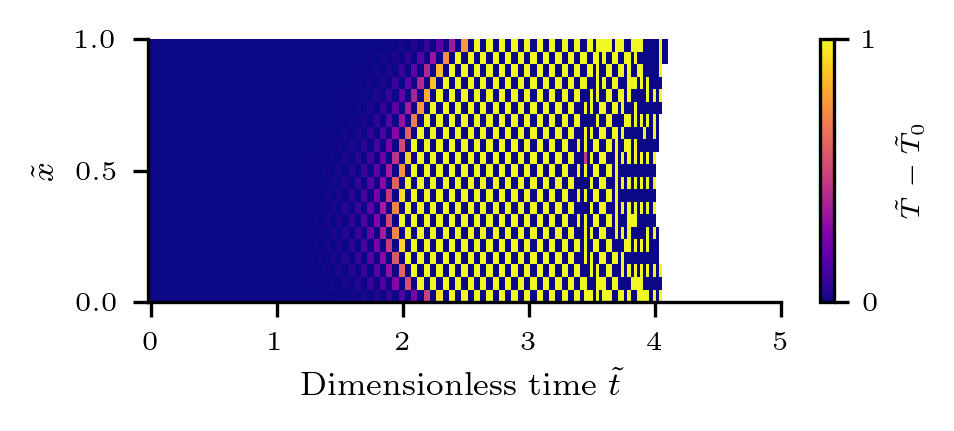}
 \\
 \hspace{6. em}\includegraphics[height=42ex]{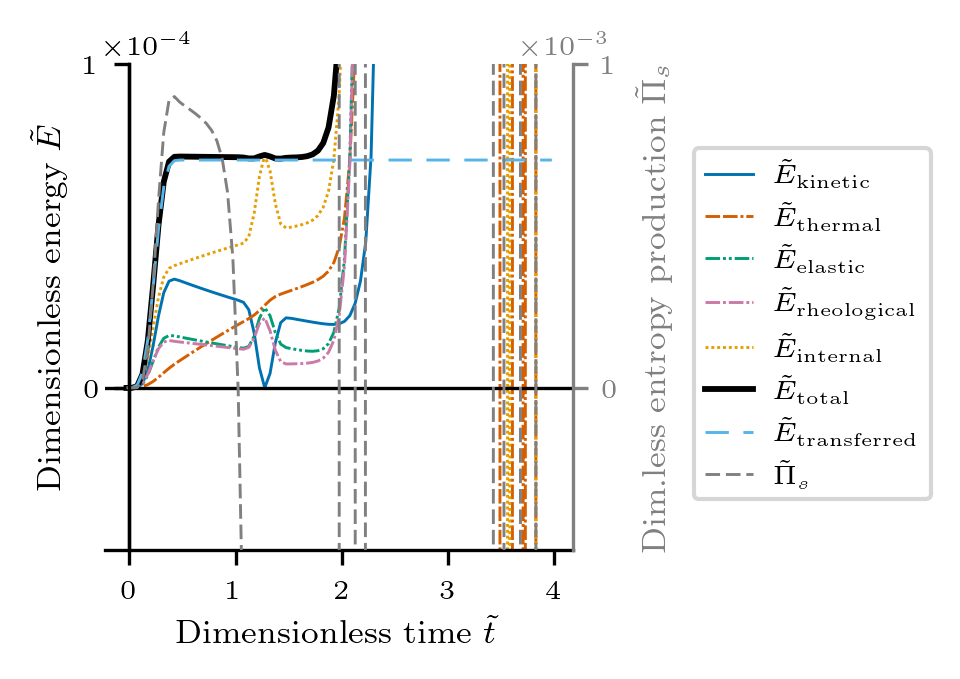}
 \caption{
Loss of stability, and how discretized entropy production rate becoming
negative reveals the problem early, well before energies become problematic.
Calculation with material C
\1 1 {see Table~\ref{tab:nondim_material_parameters}}.
 }  \label{unstable_lambda}
 \end{figure}

It may be difficult
to cope with both requirements. Figure~\ref{unstable_lambda} shows such a
case where we drove the simulation somewhat above the heat conduction related
stability criterion, which was found to be \m { \frac{\lambda}{\varrho \cex}
\frac{\Dt}{\Dx^2} \le 0.333 } via experimenting. As a generalization of the idea
introduced in \cite{pozsar2020four}, we discretize \m { \pis } in a
deliberately non-positive definite form---although, at the continuum level,
the equations would ensure its positive definiteness---:
 \begin{align}  \label{@34972}
     \9 1 { \pis }^\jpp_n & = \frac{1}{2} \frac{1}{\Dx}
 \9 2 {
\9 1 { \frac{1}{T^j_\npp} - \frac{1}{T^j_n} } \9 1 { \je }^\jpp_\np +
\9 1 { \frac{1}{T^j_n} - \frac{1}{T^j_\nmm} } \9 1 { \je }^\jpp_\nm
 } 
\nonumber \\[1. ex]
& +  \frac{1}{\hat{I}} \frac{1}{T^j_n} \sigmahat^j_n \9 2 {
\hat{I} L^\jp_n - \frac{\tau}{\Dt} \9 1 { \sigmahat^\jpp_n - \sigmahat^j_n } } .
 \end{align}
This way we make it possible that the discretized \m { \pis } takes negative
value. Since the second law of thermodynamics is related to Lyapunov
asymptotic stability, its violation at the discrete level is expected to
reveal numerical loss of stability. With an adequately chosen discretization,
entropy production rate may signal instability much before the problem
manifests itself in other quantities.
 In the example visible in Figure~\ref{unstable_lambda}, instability becomes
apparent at dimensionless time \m { \tilde t \approx 2 } while the space
integral \m { \tilde{\Pi}_s } of \m { \tilde{\pi}_s } becomes negative
already at \m { \tilde t \approx 1 }. The idea introduced in
\cite{pozsar2020four} works in the present generalized context, too: entropy
production rate proves to be a good instability indicator.

\section{Discussion}

The proposed explicit finite-difference scheme has second-order accuracy in
both space and time, has maintained the good energy-preserving property and
large-time reliability of the symplectic scheme of the reversible core of the
system, and can avoid dispersion and dissipation error. Regarding monitoring
the solution, time dependence of the various energy terms proves to make
dispersion error visible, and entropy production rate can be utilized to
point out instability at an early stage. These make the scheme applicable for
simulating viscoelastic wave propagation coupled to heat conduction via
thermal expansion, where dissipation due viscoelasticity and heat conduction
%is easy to miscalculate
 could be miscalculated
due to dissipation error, wave-based oscillations
%are easily
 could be
miscalculated due to dispersion error, and numerical settings might lead to
unstable solution.

Despite the somewhat complex structure of the scheme
\re{@34846}--\re{@37685}, it may readily be generalized to three spatial
dimensions, analogously to how \cite{pozsar2020four} generalized the scheme
set out in \cite{fulop2020thermodynamical}. This would enable to treat
rectangular/cuboid geometries. Next, along the lines of
\cite{pozsar2021wave}, extension to cylindrical setups also appears
realizable. Then, based on the experience obtained here, one becomes able to
simulate wave propagation in cylindrical rock samples for applications like
the dilatational resonant frequency method, to evaluate measurements of the
Anelastic Strain Recovery method with a higher precision, and to determine
gravitational consequences of waves around underground tunnels of
gravitational wave detectors. Other areas include simulations for plastics,
biological samples, and other solid materials where viscoelasticity is
remarkable and thermal effects also play a considerable role.
Some of these (\textit{e.g.,} \cite{huang2022peridynamic}) would require the
large-strain generalization of the current framework, which may be carried out
based on the spacetime-oriented approach
\cite{fulop2012kinematic}.
%,van2017galilean}.
%Among these is one our future aims, the large-strain generalization, to be
%able to tackle problems like discussed in \cite{huang2022peridynamic}.
The scheme communicated here is expected to be generalizable to incorporate
beyond-Fourier heat conduction (see, \textit{e.g.}, the recent review
\cite{shomali2022lagging}) as well, where calculation of the additional
irreversibility-related degrees of freedom
% poses numerical challenge that
may be successfully addressed via the scheme-building strategy presented
here -- steps in this direction have been reported in
\cite{rieth2018implicit,kovacs2020numerical}.

\subsection*{Author contributions}
D.M.T.: numerical simulations, figures, manuscript text. \'{A}.P.: numerical simulations. T.F.: conceptual idea, manuscript text. All authors have read and agreed to the published version of the manuscript.

\subsection*{Acknowledgements}

The research reported in this paper is part of project no.\ BME-NVA-02,
implemented with the support provided by the Ministry of Innovation and
Technology of Hungary from the National Research, Development and Innovation
Fund, financed under the TKP2021 funding scheme.  %% TKP2021
This work has also been supported by
 the National Research, Development and Innovation Office -- NKFIH
%KH 130378  %% KR 2018, KH_18 and
grant FK 134277, %% KR OTKA-ja, 2020-12-01- and
 the R\&D project NO.\ 2018-1.1.2-KFI-2018-00207,  %% IASR
and 
%Supported by
 the \'UNKP-22-3-I-BME-96 New National Excellence Program of the Ministry for
Culture and Innovation from the source of the National Research, Development
and Innovation Fund.

\subsection*{Conflicts of interest}
The authors have no competing interests to declare that are relevant to the content of this article.

\bibliographystyle{unsrt}
\bibliography{bibs_merged}

\end{document}